\documentclass[nofootinbib,aps,prl,reprint,
superscriptaddress]{revtex4-2}

\newif\ifprintauthors
\printauthorstrue

\usepackage[dvipsnames]{xcolor}

\usepackage{orcidlink}
\usepackage{amsmath}
\usepackage{graphicx}
\usepackage{color}
\usepackage[normalem]{ulem}

\usepackage{siunitx}
\DeclareSIUnit\parsec{pc}
\DeclareSIUnit\Mpc{\mega\parsec}

\usepackage{txfonts}
\newlength{\capheight}
\definecolor{dodgerblue}{HTML}{1E90FF}
\definecolor{rsred}{HTML}{BA0C2F}
\newcommand{\citeme}[1]{\textcolor{rsred}{CITE}}

\newcommand{\soft}[1]{\texttt{#1}}

\usepackage{acronym}
\newacro{BH}[BH]{black hole}
\newacro{BBH}[BBH]{binary black hole}
\newacro{GW}[GW]{gravitational-wave}
\newacro{PN}[PN]{post-Newtonian}
\newacro{GWTC-3.0}[GWTC-3.0]{third Gravitational-Wave Transient Catalog}
\newacro{GWTC-4.0}[GWTC-4.0]{fourth Gravitational-Wave Transient Catalog}
\newacro{HPD}{highest posterior density}
\newacro{SNR}[SNR]{signal-to-noise ratio}
\newacro{GR}[GR]{general relativity}
\newacro{NR}[NR]{numerical-relativity}
\newacro{IMR}[IMR]{inspiral--merger--ringdown}
\newacro{QNM}[QNM]{quasi-normal mode}
\newacro{LVK}[LVK]{LIGO-Virgo-KAGRA}

\newcommand{\eventname}{GW250114\xspace}

\usepackage{xspace}

\begin{document}

\title{Plunge--Merger--Ringdown Tests of General Relativity with \eventname}

\author{Leonardo Grimaldi\orcidlink{0009-0008-6703-3532}} \email{leonardo.grimaldi@aei.mpg.de}
\affiliation{Max Planck Institute for Gravitational Physics (Albert Einstein Institute), D-14476 Potsdam, Germany} 

\author{Elisa Maggio\orcidlink{0000-0002-1960-8185}}
\affiliation{INFN, Sezione di Roma, Piazzale Aldo Moro 2, 00185, Roma, Italy}
\affiliation{Max Planck Institute for Gravitational Physics (Albert Einstein Institute), D-14476 Potsdam, Germany}

\author{Lorenzo Pompili\orcidlink{0000-0002-0710-6778}}
\affiliation{Nottingham Centre of Gravity \& School of Mathematical Sciences, University of Nottingham,
University Park, Nottingham NG7 2RD, United Kingdom}
\affiliation{Max Planck Institute for Gravitational Physics (Albert Einstein Institute), D-14476 Potsdam, Germany}

\author{Alessandra Buonanno\orcidlink{0000-0002-5433-1409}}
\affiliation{Max Planck Institute for Gravitational Physics (Albert Einstein Institute), D-14476 Potsdam, Germany}
\affiliation{Department of Physics, University of Maryland, College Park, MD 20742, USA}

\begin{abstract}
The binary black hole signal GW250114, the clearest gravitational wave detected to date, offers a unique opportunity to test 
general relativity in the relativistic strong-gravity regime. How well does GW250114 agree with Einstein's predictions in the plunge--merger--ringdown stage?
To address this point, we constrain deviations from general relativity  across the plunge--merger--ringdown stage of spin-precessing binaries with a parametrized waveform model within the effective-one-body formalism.
We find that deviations from the peak gravitational-wave amplitude  and instantaneous frequency of the $(\ell, |m|)=(2,2)$ mode are constrained to about 10\% and 4\%, respectively, at 90\% credible level. These constraints are, respectively, two and four times more stringent than those obtained by analyzing GW150914.
We also constrain, for the first time, the instantaneous frequency of the $(\ell, |m|)=(4,4)$ mode at merger to about 6\%, and the time at which the gravitational-wave amplitude peaks to about $5~\mathrm{ms}$.
These results are the most precise tests of general relativity in the nonlinear regime to date, and can be employed to constrain extensions of Einsten's theory.
\end{abstract}

\maketitle

\textit{Introduction.} On January 14, 2025, the clearest gravitational-wave (GW) signal to date, GW250114\_082203 (hereafter GW250114)~\cite{LIGOScientific:2025rid}, was observed by the LIGO detectors~\cite{LIGOScientific:2014pky}. Its high network signal-to-noise ratio (SNR) of 76 allows for testing Einstein's general relativity (GR) with unprecedented accuracy~\cite{LIGOScientific:2025obp}.
Together with other signals in the fourth Gravitational-Wave Transient Catalog~\cite{LIGOScientific:2025slb}, which includes events up to the first part of the fourth observing run of the LIGO--Virgo--KAGRA (LVK) detectors~\cite{LIGOScientific:2014pky, VIRGO:2014yos, KAGRA:2020tym}, GW250114 underpins the current state of the art in strong-field tests of GR~\cite{LIGOScientific:2016lio, LIGOScientific:2019fpa, LIGOScientific:2020tif, LIGOScientific:2021sio}.

Black holes (BHs) in binary configurations in vacuum lose energy via emission of gravitational radiation and spiral toward each other until they merge~\cite{Pretorius:2005gq, Campanelli:2006fy, Baker:2005vv}. 
During the final orbits, the binary reaches relativistic velocities and highly dynamical gravitational fields, causing the two objects to rapidly coalesce, producing a short but intense burst of GWs that marks the merger.
The remnant, a highly distorted BH, settles down to a stationary configuration by emitting GWs in the ringdown~\cite{Regge:1957td, Zerilli:1970se, Vishveshwara:1970zz, Press:1971wr, Teukolsky:1973ha}. This stage can be modeled as a superposition of damped sinusoids whose frequencies and decay times, called quasi-normal modes (QNMs), depend solely on the remnant BH mass and spin~\cite{Carter71,Robinson:1975bv} in GR. 
Several studies in the last years have investigated possible signatures of beyond-GR physics in the dynamical and nonlinear regime~\cite{Okounkova:2019zjf,Bhagwat:2021kfa,Bonilla:2022dyt,Maggio:2022hre,Watarai:2023yky,Roy:2025gzv,Watarai:2025hsb} and the presence of QNMs in GW data~\cite{Brito:2018rfr,Carullo:2019flw,Isi:2021iql,Ghosh:2021mrv,Silva:2022srr,Gennari:2023gmx,Ma:2023vvr,Pompili:2025cdc,Berti:2025hly}.

Current analyses of GW250114~\cite{LIGOScientific:2025rid,LIGOScientific:2025obp} included ringdown tests of the remnant, constraints to parameterized deviations to the post-Newtonian (PN) coefficients in the inspiral phase (both single-parameter~\cite{Mehta:2022pcn,Agathos:2013upa,Meidam:2017dgf,Roy:2025gzv} and multiparameter tests using principal component analysis~\cite{Saleem:2021nsb,Mahapatra:2025cwk}), consistency tests between the inspiral and merger--ringdown portions of the waveform~\cite{Ghosh:2016qgn,Ghosh:2017gfp}, and residual tests~\cite{LIGOScientific:2020tif}. 
Notably, postmerger analyses of GW250114 reported the consistency of the data with the dominant quadrupolar $(\ell = |m| = 2)$ mode of a Kerr BH and its first overtone~\cite{LIGOScientific:2025rid, LIGOScientific:2025obp}.
By fitting a parameterized inspiral--merger--ringdown (IMR) waveform model~\cite{Pompili:2025cdc}, these studies have also constrained the fundamental $\ell = |m| = 4$ mode frequency for the first time~\cite{LIGOScientific:2025obp}. 

In this \textit{Letter}, we extend these analyses by constraining deviations from the GR predictions throughout the plunge--merger--ringdown stage, focusing on theory-independent tests. We employ the parametrized waveform model \texttt{pSEOBNRv5PHM} (hereafter \texttt{pSEOBNR})~\cite{Pompili:2025cdc,Toubiana:2023cwr,Maggio:2022hre,Ghosh:2021mrv,Brito:2018rfr}, a state-of-the-art multipolar, spin-precessing effective-one-body (EOB) waveform model for binary BHs (BBHs) in quasicircular orbits~\cite{Pompili:2023tna,Khalil:2023kep,vandeMeent:2023ols,Ramos-Buades:2023ehm}. \texttt{pSEOBNR} is a complete IMR waveform that introduces fractional deviations to the numerical relativity (NR)-informed input values for the peak amplitude and frequency, and the peak time itself, as well as the frequency and decay time of the fundamental QNMs in the ringdown to perform tests of GR with GWs.
Using this model, we place the most stringent constraints to date on deviations in the peak GW amplitude and instantaneous frequency of the $(\ell, |m|)=(2,2)$ mode, together with the first constraints on deviations to the instantaneous frequency of the $(\ell, |m|)=(4,4)$ mode at merger and the time at which the GW (2,2) mode amplitude peaks.
All tests yield results consistent with GR.

NR simulations of theories beyond GR and exotic compact objects, as well as gravitational self-force calculations, have explored potential signatures of new physics in the merger stage~\cite{Witek:2018dmd,Okounkova:2019zjf,Witek:2020uzz,Okounkova:2020rqw,East:2020hgw,East:2021bqk,Figueras:2021abd,Corman:2022xqg,AresteSalo:2022hua,Brady:2023dgu,Doneva:2023oww,AresteSalo:2023mmd,Lara:2024rwa,Corman:2024cdr,Evstafyeva:2024qvp,Lara:2025kzj,Corman:2025wun, Roy:2025kra}. 
Given the predictions for these corrections, results from theory-independent tests can be translated into constraints on the additional degrees of freedom of modified theories of gravity.

For tests of GR that use the full IMR signal, such as the \texttt{pSEOBNR} approach, waveform systematics could lead to false indications of deviations from GR already for SNRs of $\mathcal{O}(100)$~\cite{Pang:2018hjb, Hu:2022bji, Maggio:2022hre, Toubiana:2023cwr, Gupta:2024gun}. Given the network SNR of GW250114, particular attention is required when interpreting the results, as they may be subject to such systematic errors. In the Supplemental Material, we  perform injection studies to assess the impact of noise and waveform systematics on our results.\\

\textit{GW250114.} 
Using the IMR \texttt{NRSur7dq4} model~\cite{Varma:2019csw} for quasi-circular and spin-precessing binaries, Abac \emph{et al.}~\cite{LIGOScientific:2025rid} found that the wave morphology is consistent with a BBH with component masses $m_1=33.6^{+1.2}_{-0.8} M_\odot$ and $m_2=32.2^{+0.8}_{-1.3} M_\odot$ and dimensionless spin magnitudes $\chi_1\le 0.24$ and $\chi_2\le 0.26$ at 90\% credible intervals.
The results with other models~\cite{Pratten:2020ceb,Colleoni:2024knd,Pompili:2023tna,Khalil:2023kep,vandeMeent:2023ols,Ramos-Buades:2023ehm,Thompson:2023ase} are consistent with each other. 
The eccentricity is constrained to $e \le 0.03$ at a reference frequency of $13.33 \ \mathrm{Hz}$, using eccentric aligned-spin models~\cite{LIGOScientific:2025rid}. We confirm the eccentricity constraint by repeating the analysis using the \texttt{SEOBNRv5EHM} model~\cite{Gamboa:2024hli} (see the Supplemental Material for details), and conclude that eccentricity effects can be neglected when performing tests of GR with GW250114. 
\\

\textit{The waveform model.} \texttt{pSEOBNR}~\cite{Pompili:2025cdc,Toubiana:2023cwr,Maggio:2022hre,Ghosh:2021mrv,Brito:2018rfr} is a parametrized waveform model for spin-precessing BBHs in quasicircular orbits, built within the EOB formalism~\cite{Buonanno:1998gg, Buonanno:2000ef} and calibrated to NR simulations. 
It includes the subdominant modes $(\ell,|m|) = (2,1), (3,3), (3,2), (4,4), (4,3)$ and $(5,5)$ in addition to the dominant $(2,2)$ mode.
It extends the \texttt{SEOBNRv5PHM} model~\cite{Pompili:2023tna, Khalil:2023kep, vandeMeent:2023ols, Ramos-Buades:2023ehm} by introducing free parameters that encode potential deviations across the different stages of the coalescence. In the ringdown stage, it includes fractional deviations to the frequency and damping time of the fundamental QNMs~\cite{Brito:2018rfr, Ghosh:2021mrv}. In the plunge--merger stage, it allows for fractional deviations to the NR-calibrated merger amplitude and frequency of each waveform mode, as well as the instant at which the $(2,2)$-mode amplitude peaks~\cite{Maggio:2022hre}, which, in the rest of the manuscript, we associate to the merger. Additional parametrized corrections are incorporated in the NR-calibration parameters of the model in the inspiral stage~\cite{Pompili:2025cdc}.

The GW polarizations can be written in the observer's frame as
\begin{equation}
    h_+-ih_\times=\sum_{\ell,m} {}_{-2} Y_{\ell m}(\varphi,\iota) \ h_{\ell m}(t) \,,
\end{equation}
where ${}_{-2}Y_{\ell m}(\varphi,\iota)$ are the $-2$ spin-weighted spherical harmonics, with $\varphi$ and $\iota$ being the azimuthal and polar angles to the observer. 
The \texttt{SEOBNRv5PHM} model describes waveforms from spin-precessing binaries by a suitable time-dependent rotation of approximately aligned-spin waveforms in a co-precessing frame, which tracks the instantaneous direction of the orbital plane~\cite{Buonanno:2002fy,Schmidt:2010it,Boyle:2011gg,OShaughnessy:2011pmr,Pan:2013rra,Hannam:2013oca}.
The GW modes in the co-precessing frame are decomposed as
\begin{align}
    h_{\ell m}(t)=&h_{\ell m}(t)^\text{insp-plunge} \ \theta\left(t^{\ell m}_\text{match}-t\right)\nonumber\\
    &+h_{\ell m}(t)^\text{merger-RD} \ \theta\left(t-t^{\ell m}_\text{match}\right) \,,
\end{align}
where $\theta(t)$ is the Heaviside step function, $h_{\ell m}^\text{insp-plunge}$ and $h_{\ell m}^\text{merger-RD}$ are the inspiral--plunge and merger--ringdown parts of the waveform, respectively. The matching time $t^{\ell m}_\text{match}$ is chosen as the peak of the $(2,2)$ mode amplitude for all $(\ell, m)$ modes except $(5, 5)$, for which it is taken as the peak of the $(2, 2)$ harmonic minus $10M$~\cite{Pompili:2023tna}.\\
The peak-time of the $(2,2)$ mode, $t_\text{peak}^{22}$ is defined as
\begin{equation}
    t_\text{peak}^{22}=t_\text{ISCO}+\Delta t_\text{NR}\,,
\end{equation}
where $t_\text{ISCO}$ is the time at which $r = r_\text{ISCO}$, and $r_\text{ISCO}$ is the radius of the innermost-stable circular orbit (ISCO)~\cite{Bardeen:1972fi} of a Kerr BH with the remnant mass and spin~\cite{Jimenez-Forteza:2016oae, Hofmann:2016yih}. The parameter $\Delta t_\text{NR}$ is a free parameter calibrated to aligned-spin NR waveforms ~\cite{Pompili:2023tna}.

The inspiral--plunge waveform, $h_{\ell m}^\text{insp-plunge}$, is constructed on a factorization of the PN GW modes~\cite{Damour:2008gu, Damour:2007xr, Pan:2010hz}, evaluated on the dynamics obtained from the EOB equations of motion~\cite{Buonanno:2000ef}.
Non-quasicircular corrections enforce that the modes' amplitude and first two derivatives, and their frequency and first derivative match \textit{input values} extracted from NR simulations at $t^{\ell m}_\text{match}$.
The model has two calibration parameters in the EOB Hamiltonian, namely $a_6$, that is a 5PN parameter that enters the non-spinning potential, and $d_\text{SO}$, that is a 4.5PN spin-orbit parameter.

For all harmonics, except for $(\ell, |m|) = (3, 2)$ and $(4, 3)$, the merger--ringdown waveform employs the following Ansatz,
\begin{equation}
    h_{\ell m}^\text{merger-RD}(t)=\nu \ \tilde A_{\ell m}(t) \ e^{i\tilde\phi_{\ell m}(t)} \ e^{-i\sigma^{\mathrm{CP}}_{\ell m 0}\left(t-t^{\ell m}_\text{match}\right)} \,,
\end{equation}
where $\nu=m_1m_2/M^2$ is the symmetric mass-ratio, $M=m_1+m_2$ is the total mass, $\tilde A_{\ell m}$ and $\tilde\phi_{\ell m}$ are time-dependent amplitude and phase functions, and $\sigma_{\ell m 0}^{\mathrm{CP}}$ is the complex frequency of the least-damped QNM of the remnant BH in the co-precessing frame. 
The QNM frequencies obtained from BH perturbation theory are formally valid in the $\mathbf{J}_f$-frame, where the $z$-axis is aligned with the spin of the final BH. The $\mathbf{J}_f$-frame QNMs, $\sigma_{\ell m 0}$, are related to the co-precessing frame QNMs by Eq.~(35) of Ref.~\cite{Hamilton:2023znn}.

The real and imaginary parts of $\sigma_{\ell m 0}$ are related to the QNM oscillation frequency and damping time as follows:
\begin{equation}
    f_{\ell m 0}=\frac{1}{2\pi}\text{Re}\left(\sigma_{\ell m 0}\right)\,, \quad \tau_{\ell m 0}=-\frac{1}{\text{Im}\left(\sigma_{\ell m 0}\right)} \,.
\end{equation}
Fractional deviations to the frequency and decay time of the fundamental QNMs in the $\mathbf{J}_f$-frame are introduced as:
\begin{equation}
    f_{\ell m 0}=f_{\ell m 0}^\text{GR}\left(1+\delta f_{\ell m 0}\right)\,, \quad \tau_{\ell m 0}=\tau_{\ell m 0}^\text{GR}\left(1+\delta \tau_{\ell m 0}\right)\,,
\end{equation}
where the GR predictions for these quantities ($f_{\ell m 0}^\text{GR}$ and $\tau_{\ell m 0}^\text{GR}$) are obtained from the  mass and spin of the remnant BH, estimated using NR fits on the measured component masses and spins~\cite{Jimenez-Forteza:2016oae, Hofmann:2016yih}.

Fractional deviations to the NR-informed input values for the co-precessing frame modes' amplitude and frequency at $t=t^{\ell m}_\text{match}$ are introduced as
\begin{equation}
    |h_{\ell m}|=|h_{\ell m}^\text{NR}|\left(1+\delta A_{\ell m}\right)\,, \quad \omega_{\ell m}=\omega_{\ell m}^\text{NR}\left(1+\delta \omega_{\ell m}\right)\,,
\end{equation}
where $|h_{\ell m}^\text{NR}|$ and $\omega_{\ell m}^\text{NR}$ are calibrated to NR simulations at the matching point $t^{\ell m}_\text{match}$ (see Sec. IIIA. of Ref.~\cite{Pompili:2023tna} for details).

The model also allows for changes to $t^{\ell m}_\text{match}$ by modifying the time-shift parameter $\Delta t_\text{NR}$ as:
\begin{equation}
    \Delta t_\text{NR}\rightarrow\Delta t_\text{NR}-\delta \Delta t\,.
    \label{ddt}
\end{equation}
The deviation parameter $\delta \Delta t$ has, by definition, a lower bound. Indeed, if this parameter is too negative, $t^{\ell m}_\text{match}$ can overshoot the end of the BBH dynamics~\cite{Pompili:2023tna}.

Additive deviations to the EOB calibration parameters $a_6$ and $d_\text{SO}$
are introduced as follows:
\begin{equation}
    a_6 \rightarrow a_6+\delta a_6 \,,\quad
    d_\text{SO} \rightarrow d_\text{SO}+\delta d_\text{SO} \,.
    \label{calibration}
\end{equation}

\textit{Prior choices.} We assume prior distributions on the GR parameters consistent with Ref.~\cite{LIGOScientific:2025yae}. 
Specifically, we sample the masses using the chirp mass ($\mathcal{M}=\nu^{3/5}M$) and mass ratio ($q=m_2/m_1$), with uniform priors in component masses. The priors on the dimensionless spin vectors are uniform in magnitude $\chi_i \in [0,0.99]$, and isotropically distributed in the unit sphere for the spin directions. For the distance, we employ a
uniform prior in the comoving-frame of the source.

The fractional deviations are, by definition, larger than $-1$. For them, we assume uniform priors with a lower limit set to be $-0.8$, to avoid anomalies near $-1$, and an upper limit of $1$, with the only exception of $\delta A_{44}$ which has an upper limit extended to $4$. The time shift deviation $\delta \Delta t$, instead, has a lower limit incorporated in the \texttt{pSEOBNR} model to ensure that $t^{\ell m}_\text{match}$ is not after the last point of the BBH dynamics. The upper limit on its uniform prior is set to $50M$. For $\delta a_6$ and $\delta d_\text{SO}$, we set a uniform prior between $-100$ and $100$.

We use \texttt{Bilby}~\cite{Ashton:2018jfp, Romero-Shaw:2020owr} with the \texttt{dynesty} nested sampler~\cite{Speagle:2019ivv} to sample the posterior distributions of the GR and deviation parameters.\\

\begin{figure}
    \centering
    \includegraphics[width=\linewidth]{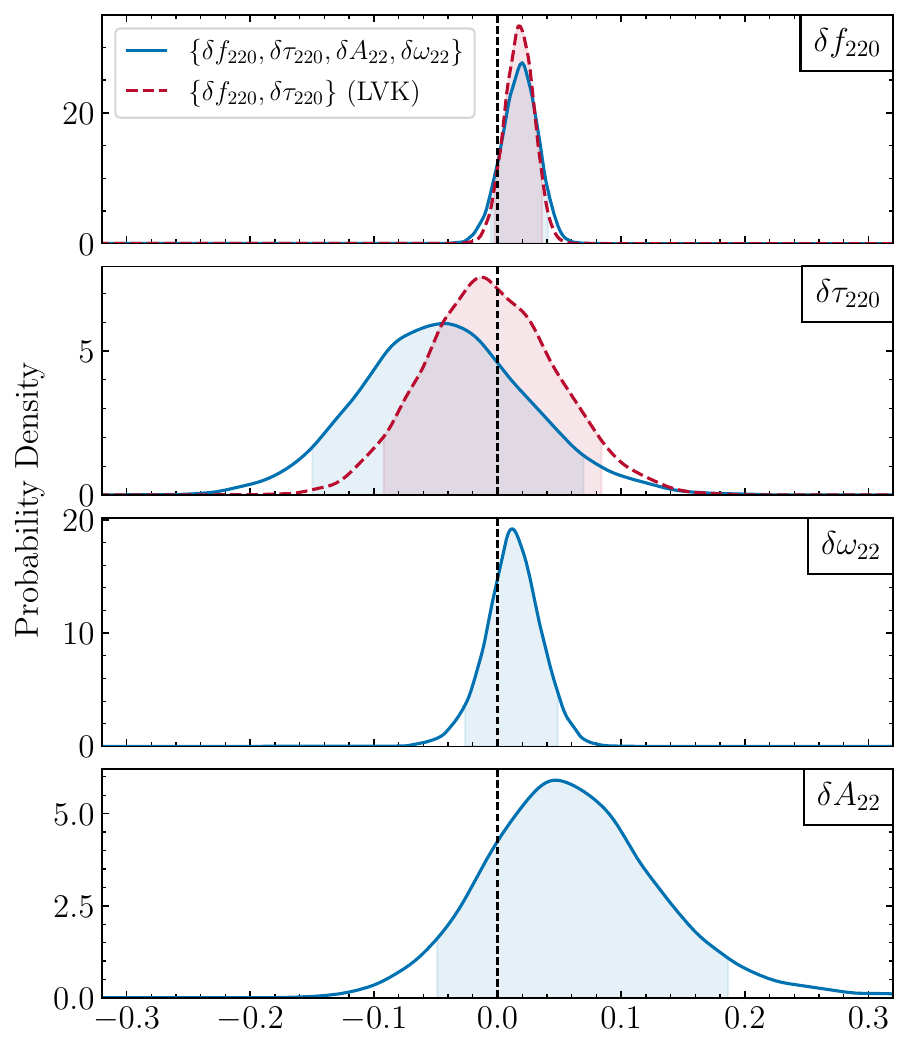}
    \caption{The one-dimensional posterior distributions on the merger--ringdown deviation parameters for our analysis (blue lines) and the LVK analysis~\cite{LIGOScientific:2025obp} (red lines), which included only ringdown deviations. Both analyses were performed using the \texttt{pSEOBNRv5PHM} waveform model where the shaded areas indicate 90\% credible intervals. The vertical lines mark the null GR expectation. The inferred values of $\delta A_{22}$ and $\delta\omega_{22}$ are consistent with GR.}
    \label{posteriors1}
\end{figure}

\textit{The GW250114 analysis.} We first analyze GW250114 by allowing for deviations to the merger amplitude and frequency of the dominant $(2,2)$ mode, as well as the ringdown $(2,2,0)$ QNM. The results are shown in Fig.~\ref{posteriors1}, where the frequency and damping time of the $(2,2,0)$ QNM are consistent with the LVK analysis in Ref.~\cite{LIGOScientific:2025obp}. The merger amplitude and frequency of the $(2,2)$ mode are consistent with the predictions of GR and are exceptionally well constrained, with $\delta \omega_{22}=0.01^{+0.04}_{-0.04}$ and $\delta A_{22}=0.06^{+0.13}_{-0.11}$ at $90\%$ credible intervals. The constraint on the merger amplitude is roughly twice as stringent as the one obtained from the analysis of GW150914~\cite{LIGOScientific:2016aoc,Maggio:2022hre}, while the constraint on the merger frequency is about four times tighter. For comparison, the GW150914 analysis yielded $\delta \omega_{22}=0.00^{+0.17}_{-0.12}$ and $\delta A_{22}=-0.01^{+0.27}_{-0.19}$. The remarkable improvement of the GR constraints at merger is explained by the larger SNR of the event, of a factor of three with respect to GW150914.

When interpreting our inference on these parameters, it is important to note that in the \texttt{SEOBNRv5PHM} model they are calibrated to NR simulations of aligned-spin binaries. The model extends these fits to spin-precessing systems by evaluating them for the spin components perpendicular to the orbital plane at the matching time.
Because GW250114 is consistent with a binary with negligible spin precession, the error introduced by approximating the input values with approximately aligned-spin fits is subdominant relative to the statistical uncertainties. We compare the merger amplitude and frequency predicted by \texttt{SEOBNRv5PHM} and \texttt{NRSur7dq4} for the posterior samples of GW250114 with \texttt{NRSur7dq4}, finding average differences of around $1\%$, smaller than the statistical uncertainty of our analysis ($\gtrsim 4\%$).

As a further step, we extend the analysis of GW250114 to probe the $(4,4)$ mode. We perform analyses including only the merger parameters and both merger and ringdown parameter deviations. We find that their results agree with each other, and the results for the $(2,2,0)$ and $(4,4,0)$ QNMs are compatible with those of the LVK analysis~\cite{LIGOScientific:2025obp}. Quite notably, this shows that the constraints on the $(4,4,0)$ QNM are a robust result, since they do not change when allowing for the merger amplitude and frequency to deviate from their GR predictions (see the Supplemental Material for a more detailed discussion).

\begin{figure}
    \centering
    \includegraphics[width=\linewidth]{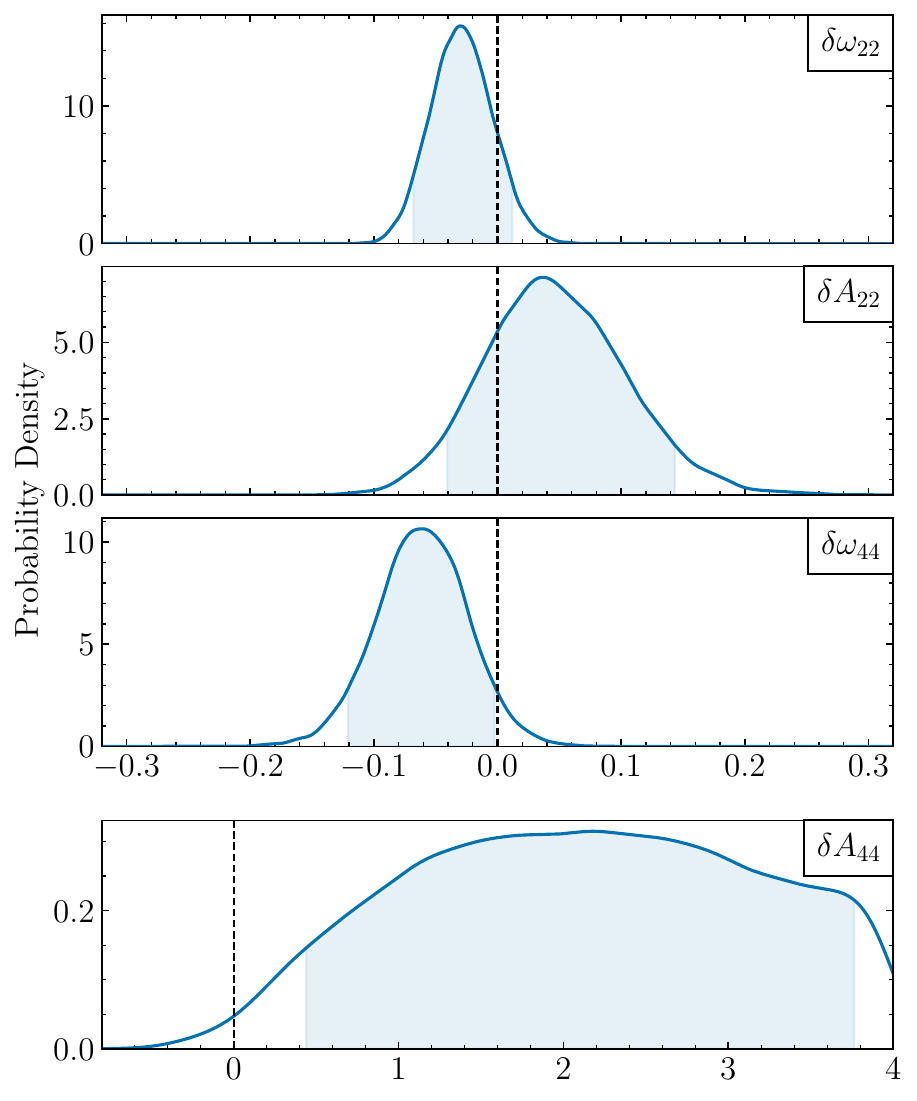}
    \caption{The one-dimensional posterior distributions on the merger parameters for GW250114, where the shaded areas indicate 90\% credible intervals. The vertical lines mark the null GR expectation. The inferred values of $\delta A_{22}$ and $\delta \omega_{22}$ are consistent with GR, while $\delta \omega_{44}$ is marginally consistent with GR. The posterior distribution on $\delta A_{44}$ is unconstrained and rails against the upper bound (note the different scale with respect to the other parameters).}
    \label{posteriors2}
\end{figure}

Since the analyses agree with each other, in the following we present the one with deviations to the merger parameters $\delta A_{22}, \delta \omega_{22},\delta A_{44}$ and $\delta \omega_{44}$.
The results are summarized in Fig.~\ref{posteriors2}. Remarkably, we constrain, for the first time, the merger frequency of the subdominant $(4,4)$ mode, obtaining $\delta \omega_{44}=-0.06^{+0.06}_{-0.06}$, which is marginally compatible with GR. The amplitude of this mode remains unconstrained.

To better understand why the posterior distribution of $\delta A_{44}$ remains unconstrained and rails against the upper bound of the prior range, we perform several injection studies (see the Supplemental Material for details). First, we find a correlation between the posterior samples with large $\delta A_{44}$ and the inclination angle, $\iota$. This causes the posterior distribution on $\iota$ to shift away from the GR parameter estimation (see Fig.~\ref{correlations} in the Supplemental Material). Since the inclination angle correlates with the luminosity distance, $D_L$, we also find a bias with the GR recovery for this parameter. 
We perform zero-noise injections with \texttt{SEOBNRv5PHM} and \texttt{NRSur7dq4} waveforms, and recover $\delta A_{44}$ with wide and uninformative posteriors that no longer rail against the upper limit of the prior (see Fig.~\ref{44} in the Supplemental Material). This result excludes the possibility that waveform systematics can be responsible for the railing of the $\delta A_{44}$ posterior.
We then inject the \texttt{SEOBNRv5PHM} waveform in ten different Gaussian-noise realizations, and find that one of them can induce the railing observed in the real data (see the thick ocher line in Fig.~\ref{44}). We conclude that the correlation between $\delta A_{44}$ and $\iota$ causes the posterior on the first parameter to have a long tail, while the specific noise realization of the event produces a shift away from GR, causing it to rail against the upper bound. The interplay between the low SNR of the $(4,4)$ mode, the correlation with the inclination angle, and the noise make the measurement of $\delta A_{44}$ particularly difficult for this specific event.

We also notice the presence of a mild correlation between $\delta\omega_{22}$ and $\delta\omega_{44}$. Indeed, the posterior distribution on $\delta \omega_{22}$ is shifted towards negative values with respect to the analysis with merger deviations only to the $(2,2)$ mode. 
By injecting \texttt{SEOBNRv5PHM} and \texttt{NRSur7dq4} waveforms in zero noise, we find that the correlation between $\delta\omega_{22}$ and $\delta\omega_{44}$ is already present.  The introduction of the Gaussian noise can accentuate the shifts on $\delta\omega_{44}$, making it marginally compatible with GR. 
We conclude that the correlation between $\delta \omega_{22}$ and $\delta \omega_{44}$ is genuine and does not depend on waveform systematics. The shift of $\delta\omega_{44}$ is instead explained by the interplay between the correlation with $\delta\omega_{22}$ and the specific noise realization of the event.
More details are presented in the Supplemental Material. 

\begin{figure}
    \centering
    \includegraphics[width=\linewidth]{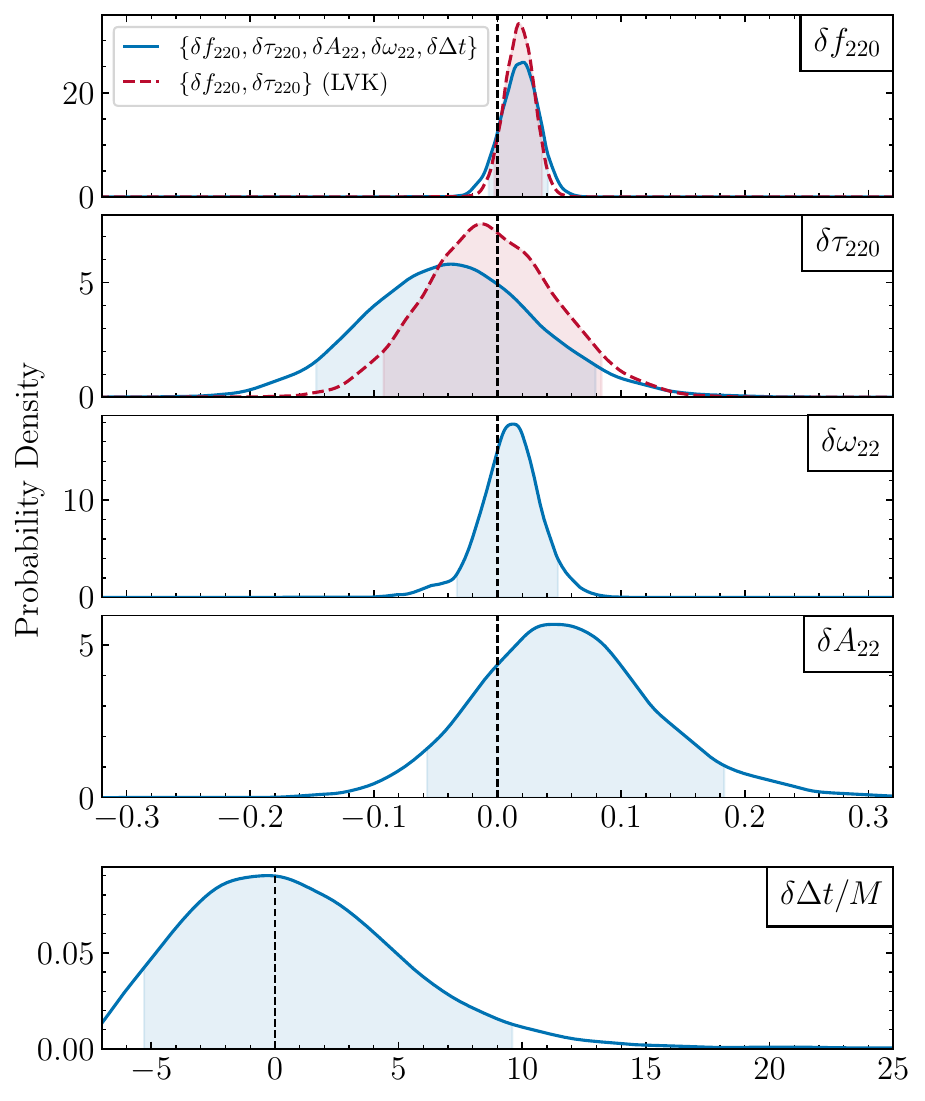}
    \caption{The one-dimensional posterior distributions on the merger--ringdown deviation parameters for our analysis (blue lines) and the LVK analysis~\cite{LIGOScientific:2025obp} (red lines), which included only ringdown deviations. Both analyses were performed using the \texttt{pSEOBNRv5PHM} waveform model. The shaded areas indicate 90\% credible intervals. The vertical lines mark the null GR expectation. We constrain for the first time the merger time shift to be consistent with GR.}
    \label{posteriors3}
\end{figure}

Finally, we perform an analysis of GW250114 allowing for deviations in the merger amplitude, frequency and peak time of the $(2,2)$ mode, as well as the $(2,2,0)$ QNM. The results are summarized in Fig.~\ref{posteriors3}. We constrain, for the first time, the merger time shift to $\delta \Delta t=0.5^{+9.1}_{-5.8}M$. The $90\%$ credible interval corresponds to about $5$ ms, to be compared with the signal length of about $550$ ms. This finding is consistent with the injection studies in Ref.~\cite{Maggio:2022hre} with $\text{SNR} \sim 100$  (note that a different parametrization on the time shift parameter was used).

We also analysed GW250114 by allowing for parametrized deviations to the inspiral calibration parameters $a_6$ and $d_\text{SO}$, however the analyses returned uninformative posteriors. This result can be explained by the fact that $d_\text{SO}$ has a prefactor proportional to the effective spin $\chi_\text{eff}$ in the EOB Hamiltonian. Since for this event $\chi_\text{eff}$ is well measured to be small, $d_\text{SO}$ is hard to constrain. Moreover $a_6$, which is a 5PN correction to the EOB Hamiltonian, likely needs higher SNR and longer signals to be constrained.

Overall, the stringent \texttt{pSEOBNR} results can improve current constraints on gravity theories beyond GR and exotic compact objects, provided model predictions for beyond-GR modifications to merger observables are available.
While accurate merger amplitudes and frequencies require NR calibration, and sufficiently accurate NR simulations for beyond-GR theories are not yet available, we can nonetheless perform an illustrative mapping for the time at which the GW amplitude peaks.
As an example, we map the observational bound of $\delta\Delta t$ to Einstein--dilaton--Gauss--Bonnet (EdGB) gravity using the \soft{SEOBNR-EdGB} model of Ref.~\cite{Julie:2024fwy}. In this model, the merger time shifts because EdGB corrections modify the ISCO location, generally leading to an earlier merger than in GR. One can therefore determine the largest coupling compatible with the inferred bound from the analysis of GW250114.
Using the maximum-likelihood \texttt{NRSur7dq4} parameters, we find that a coupling of $\sqrt{\alpha_{\rm GB}}\simeq5.8 \ \mathrm{km}$ yields a peak-time shift of $\sim 9M$ relative to GR at the upper edge of the 90\% credible interval. Although this constraint is weaker than the tightest existing bounds (i.e., $\sqrt{\alpha_{\rm GB}}\lesssim0.3 \ \mathrm{km}$~\cite{Julie:2024fwy, Sanger:2024axs}), largely because the source is near equal-mass and relatively massive, it illustrates how agnostic merger-deviation parameters can be approximately mapped to specific beyond-GR theories. A fully robust interpretation of this mapping would ultimately require NR-calibrated predictions for this parameter, as done in GR. \\

\textit{Conclusions.} We tested GR in the plunge--merger--ringdown stage of GW250114, the clearest event to date.
We found that the $(2,2)$ mode peak amplitude and instantaneous frequency are constrained to about 10\% and 4\%, respectively, at 90\% credible level. These results are, respectively, two and four times more stringent than the ones obtained for GW150914. We also constrained, for the first time, the instantaneous frequency of the $(4,4)$ mode at merger to about 6\%, and the time at which the GW $(2,2)$ mode amplitude peaks to about 5 ms. Thanks to the high SNR of this event, these results offer the most accurate tests of GR in the nonlinear regime to date and mark a significant improvement over previous analyses, heralding a new era of precision tests of GR.

We performed injection studies in zero and Gaussian noise to assess the correlations between the binary parameters and the parametrized deviations from GR. Our results confirm that waveform systematics does not affect the GW250114 analyses, while the specific noise realization of the event can slightly affect the parameter estimation.

The analyses did not constrain the $(4,4)$ mode amplitude at merger, as the posterior distribution on this parameter is wide and rails against the upper bound of its prior range. This behavior can be explained by the correlation with the inclination angle and the specific noise realization of the event, which make the measurement of $\delta A_{44}$ particularly difficult in this specific case. The lack of a meaningful constraint on deviations in the $(4,4)$ mode amplitude at merger is consistent with the findings of Refs.~\cite{Gupta:2025paz, Chandra:2025jfc}, which considered a global deviation to the $(4,4)$-mode amplitude and similarly found no significant constraint for GW250114, with the posterior showing little support near zero.

We have shown that the constraints on the $(2,2,0)$ and $(4,4,0)$ QNMs do not depend on the assumptions that are made for the amplitudes and the exact point at which inspiral and merger--ringdown are attached. These results provide evidence for the robustness of the QNM constraints found by the LVK collaboration~\cite{LIGOScientific:2025obp}.

While finalizing the paper, we became aware of an independent study, Chandra \textit{et al.},~\cite{Chandra:2025jfc}, that constrains the GW250114 merger’s amplitude of the $(2,2)$ mode using \texttt{pTEOBResumS-Dalí}.\\

\textit{Acknowledgements.} 
We thank Anuradha Gupta for comments on the manuscript.
L.G. and E.M. acknowledge funding from the Deutsche Forschungsgemeinschaft (DFG): Project No. 386119226.
E.M. is supported by the European Union’s Horizon Europe research and innovation programme under the Marie Skłodowska-Curie grant agreement No. 101107586. 
L.P. is supported by a UKRI Future Leaders Fellowship (grant number MR/Y018060/1). 
We also acknowledge the computational resources provided
by the Max Planck Institute for Gravitational Physics
(Albert Einstein Institute), Potsdam, in particular, the
Hypatia cluster.
The material presented in this paper is based upon
work supported by National Science Foundation’s (NSF)
LIGO Laboratory, which is a major facility fully funded
by the NSF. This research has made use of data or software obtained from the Gravitational Wave Open Science
Center (gwosc.org), a service of LIGO Laboratory, the
LIGO Scientific Collaboration, the Virgo Collaboration,
and KAGRA. LIGO Laboratory and Advanced LIGO are
funded by the United States National Science Foundation
(NSF) as well as the Science and Technology Facilities
Council (STFC) of the United Kingdom, the Max-Planck-Society (MPS), and the State of Niedersachsen/Germany
for support of the construction of Advanced LIGO and
construction and operation of the GEO600 detector. Additional support for Advanced LIGO was provided by the
Australian Research Council. Virgo is funded, through
the European Gravitational Observatory (EGO), by the
French Centre National de Recherche Scientifique (CNRS),
the Italian Istituto Nazionale di Fisica Nucleare (INFN)
and the Dutch Nikhef, with contributions by institutions
from Belgium, Germany, Greece, Hungary, Ireland, Japan,
Monaco, Poland, Portugal, Spain. KAGRA is supported
by Ministry of Education, Culture, Sports, Science and
Technology (MEXT), Japan Society for the Promotion
of Science (JSPS) in Japan; National Research Foundation (NRF) and Ministry of Science and ICT (MSIT) in
Korea; Academia Sinica (AS) and National Science and
Technology Council (NSTC) in Taiwan.

%%%%%%%%%%%%%%%%%%%%%%%%
\bibliography{References}
%%%%%%%%%%%%%%%%%%%%%%%%

%%%Supplemental Material%%%
\newpage
\onecolumngrid
\setlength{\parindent}{15pt}

\newpage

\section*{SUPPLEMENTAL MATERIAL}

\subsection*{Eccentricity constraint using \texttt{SEOBNRv5EHM}}

\begin{figure}[htb]
    \centering
    \includegraphics[width=0.45\linewidth]{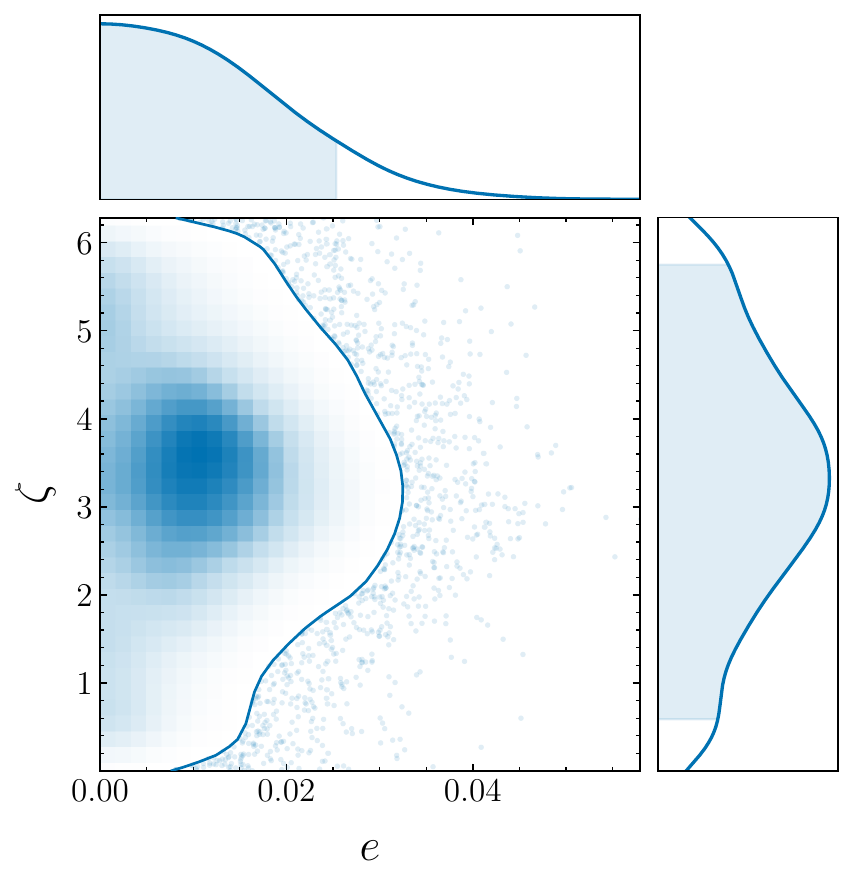}
    \caption{The one- and two-dimensional posterior distributions on the orbital eccentricity $e$ and the relativistic anomaly $\zeta$ obtained with the aligned-spin, eccentric \texttt{SEOBNRv5EHM} model~\cite{Gamboa:2024hli}. All contours indicate 90\% credible regions. The eccentricity is constrained to $e < 0.025$ at 90\% confidence, consistent with the bound $e < 0.03$ reported by the LVK collaboration~\cite{LIGOScientific:2025rid}.}
    \label{ecc}
\end{figure}

Unmodeled orbital eccentricity can lead to biased parameter estimates and apparent deviations from tests of GR \cite{Saini:2022igm,Bhat:2022amc,Narayan:2023vhm,Shaikh:2024wyn}. Because the \texttt{pSEOBNR} waveform model neglects orbital eccentricity, we seek to confirm that this assumption is valid for GW250114.
To do so, we perform an analysis using the aligned-spin, eccentric \texttt{SEOBNRv5EHM} model~\cite{Gamboa:2024hli}. We adopt the same settings and priors as in the main text, but restricting the spins to be aligned with the orbital angular momentum. We additionally sample over the orbital eccentricity and the relativistic anomaly, both defined at the orbit-averaged reference frequency of $13.33~\mathrm{Hz}$, assuming uniform priors. The posterior distribution for eccentricity and relativistic anomaly are shown in Fig.~\ref{ecc}.
The results are consistent with the quasi-circular assumption: the eccentricity is constrained to $e < 0.025$ at 90\% confidence.
These findings agree with the bound $e < 0.03$ reported by the LVK collaboration~\cite{LIGOScientific:2025rid}, obtained using a combination of the \texttt{SEOBNRv5EHM} and \texttt{TEOBResumS-Dalí}~\cite{Nagar:2024dzj} waveforms.

\subsection*{Robustness of the LVK analysis on the $(4,4,0)$ QNM against the addition of  merger parameters}

\begin{figure}
    \centering
    \includegraphics[width=0.45\linewidth]{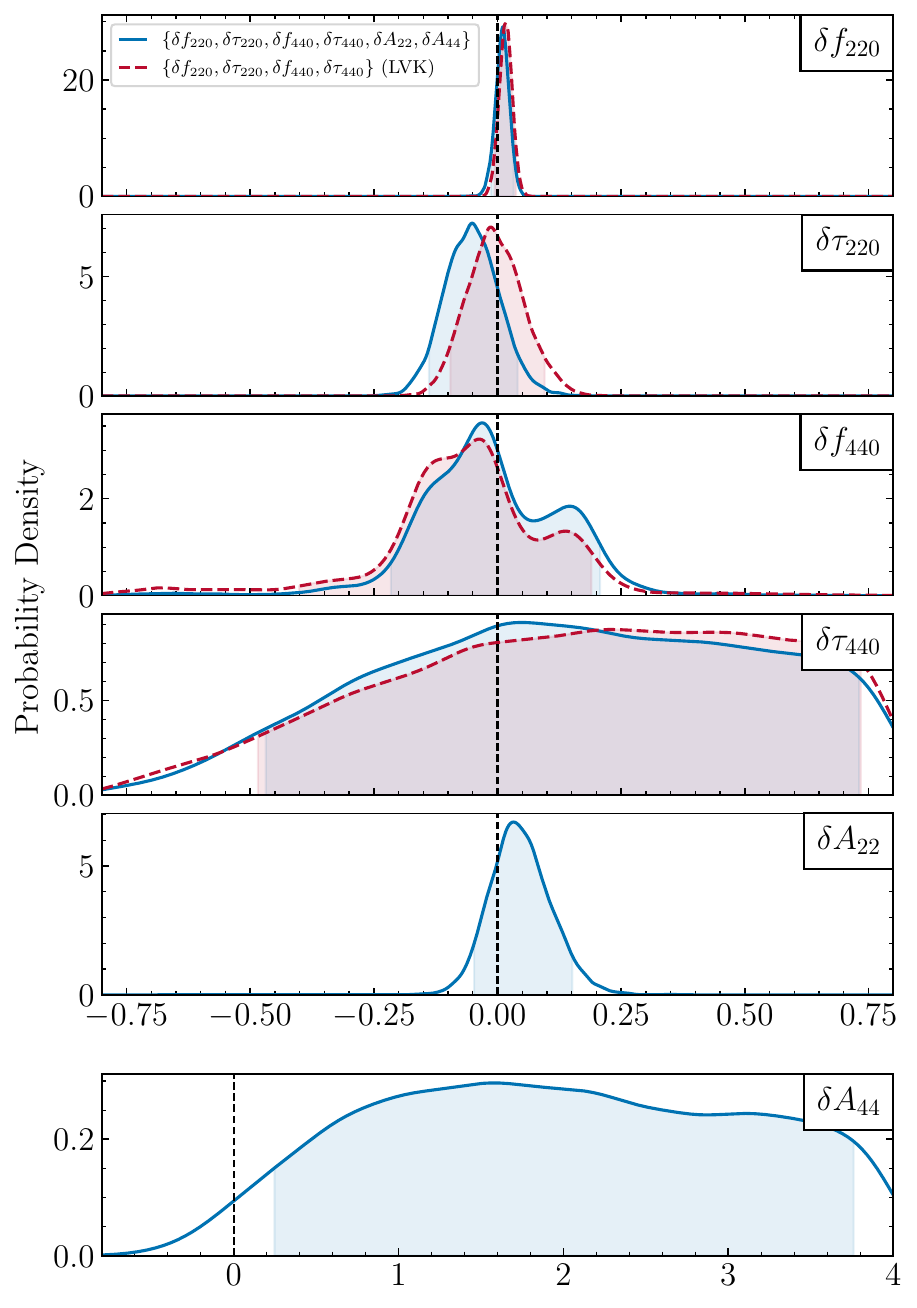}
    \caption{The one-dimensional posterior distributions on the merger--ringdown deviation parameters for our analysis (blue lines) and the LVK analysis~\cite{LIGOScientific:2025obp} (red lines), which considered only ringdown deviations. Both analyses were performed using the \texttt{pSEOBNRv5PHM} waveform model. The shaded areas indicate 90\% credible intervals. The vertical lines mark the null GR expectation. The posterior distributions on the $(2,2,0)$ and $(4,4,0)$ QNM deviations are almost unchanged when the merger parameters are added to the analysis, providing evidence on the robustness of the LVK analysis on the $(4,4,0)$ QNM.}
    \label{amplitudes}
\end{figure}

When performing the analyses including the $(2,2,0)$ and the $(4,4,0)$ QNMs deviations together with the merger amplitude and frequency deviations to the $(2,2)$ and $(4,4)$ modes, the constraints on the $(4,4,0)$ QNM deviations do not change significantly with respect to the analysis performed by the LVK collaboration~\cite{LIGOScientific:2025rid}, which considered only the ringdown deviations. Figure~\ref{amplitudes} shows that the constraints on the $(4,4,0)$ QNM do not strongly depend on the assumptions made for the mode amplitudes and the precise attachment point between the inspiral and merger--ringdown waveforms. Overall, these results provide further evidence for the robustness of the QNM constraints reported by the LVK collaboration~\cite{LIGOScientific:2025obp}.

\subsection*{Correlation between the merger frequency deviations of the $\boldsymbol{(2,2)}$ and $\boldsymbol{(4,4)}$ modes}

\begin{figure}[htb]
    \centering
    \includegraphics[width=0.45\linewidth]{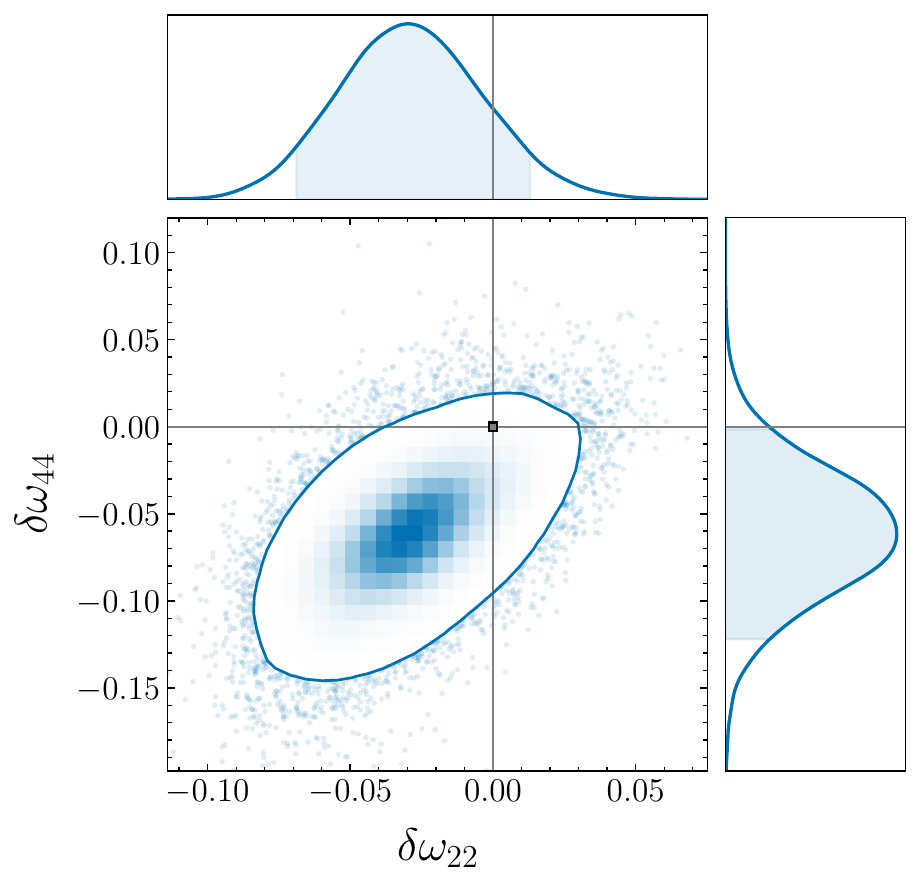}
    \caption{The one- and two-dimensional posterior distributions on the merger frequency deviations for the $(2,2)$ and $(4,4)$ mode in the real data analysis. All contours indicate 90\% credible regions. The vertical and horizontal lines mark the null GR expectation.}
    \label{dw22_dw44}
\end{figure}

\begin{figure*}[htb]
    \centering
    \includegraphics[width=0.4\linewidth]{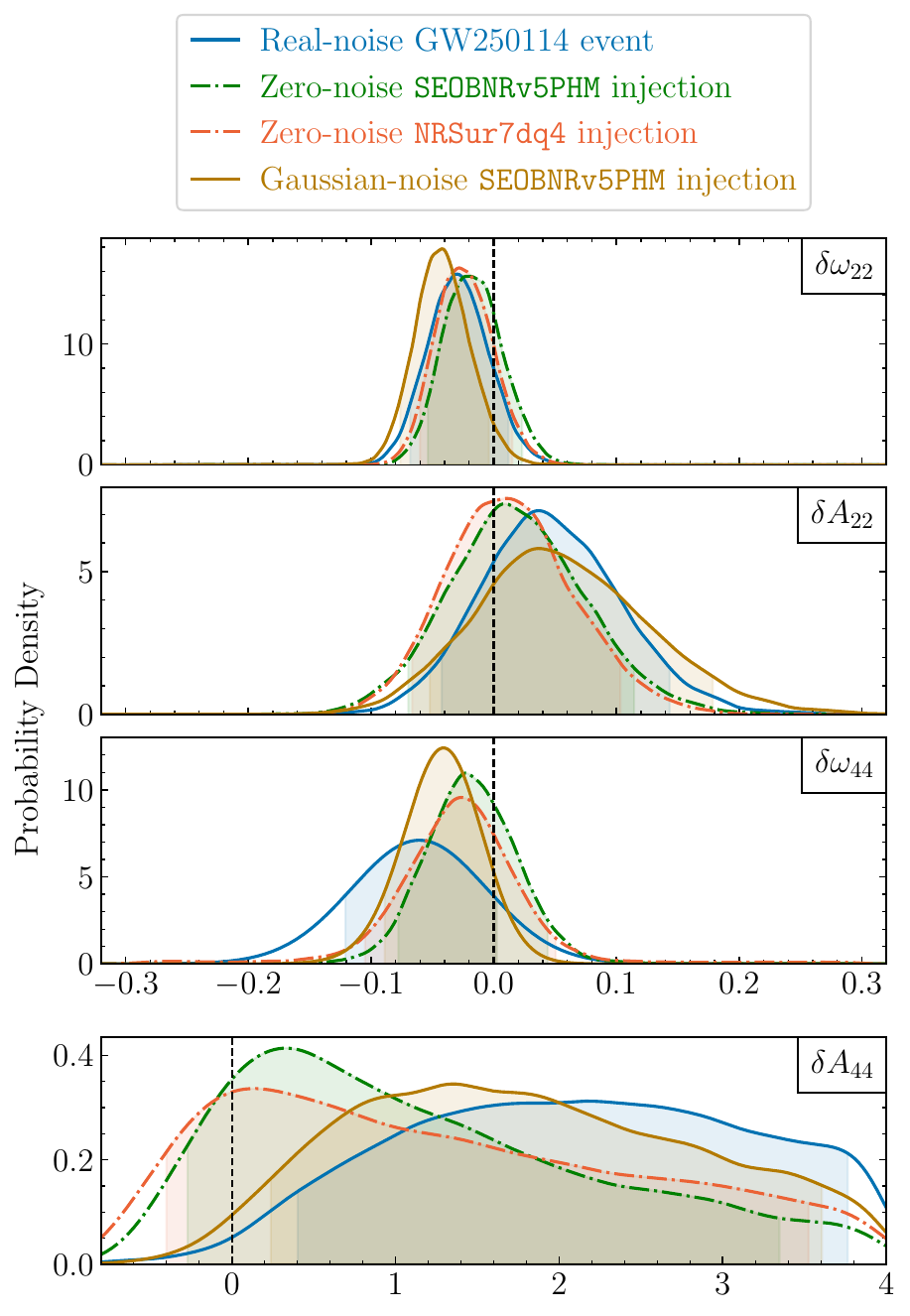}
    \caption{The one-dimensional posterior distributions on the merger parameters for the \texttt{pSEOBNR} analysis of GW250114 with $\{\delta A_{22}, \delta \omega_{22}, \delta A_{44}, \delta \omega_{44}\}$ (blue curve) compared to the posterior distributions on the merger parameters for the same analysis with an \texttt{SEOBNRv5PHM} injection in zero-noise (green curve), \texttt{\texttt{NRSur7dq4}} injection in zero noise (red curve), and one \texttt{SEOBNRv5PHM} injection in Gaussian noise (ocher curve). The shaded areas indicate 90\% credible intervals. The vertical lines mark the null GR expectation.}
    \label{inj}
\end{figure*}

The deviation to the merger frequency of the $(2,2)$ mode  correlates with the one of the $(4,4)$ mode. Fig.~\ref{dw22_dw44} shows a corner plot for the posterior distributions on $\delta \omega_{22}$ and $\delta \omega_{44}$  from the analysis of GW250114 including only the merger parameters.  
The posterior distribution on $\delta \omega_{22}$ is shifted towards negative values with respect to the analysis with merger deviations only to the $(2,2)$ mode shown in Fig.~\ref{posteriors1}. Moreover, the posterior distribution of $\delta \omega_{44}$ is marginally compatible with GR. To better understand this result, we perform a number of synthetic-signal injection studies. In particular, we use \texttt{SEOBNRv5PHM} to generate a mock signal with the maximum likelihood values of the GW250114 analysis, inject it into zero noise and recover it using the \texttt{pSEOBNRv5PHM} model with $\{\delta A_{22}, \delta\omega_{22}, \delta A_{44}, \delta\omega_{44}\}$ as deviation parameters. The results of this analysis are shown in Fig.~\ref{inj} (green curves). All 90\% confidence intervals on the merger parameters are consistent with GR, marked with dashed vertical lines.

When we inject the same synthetic signal in different realizations of Gaussian noise, we obtain random deviations around the distribution obtained in zero noise. In particular, one realization of noise makes the posterior distribution on $\delta \omega_{44}$ shift towards negative values at the point that it becomes marginally compatible with GR (ocher curves in Fig.~\ref{inj}). This same noise realization causes also the posterior on $\delta A_{44}$ to shift away from GR towards positive values, corroborating the hypothesis that the noise can cause this parameter to rail against the upper bound (see the dedicated section below in the Supplemental Material).

Finally, we use \texttt{NRSur7dq4} to generate a mock signal and inject it in zero noise (red curves in Fig.~\ref{inj}). The results of the analysis are very similar to the \texttt{SEOBNRv5PHM} injection in zero noise. Therefore, we conclude that waveform systematics is not affecting the analysis of GW250114 with merger deviations.

We conclude that the correlation between $\delta \omega_{22}$ and $\delta \omega_{44}$ is genuine and can explain the shift of $\delta\omega_{22}$ towards negative values. The shift of $\delta\omega_{44}$ is explained by the interplay between the correlation with $\delta\omega_{22}$ and the specific noise realization of the event.

\subsection*{Correlation between the merger amplitude deviation of the $(4,4)$ mode and the inclination angle}

\begin{figure}
    \centering
    \includegraphics[width=0.49\linewidth]{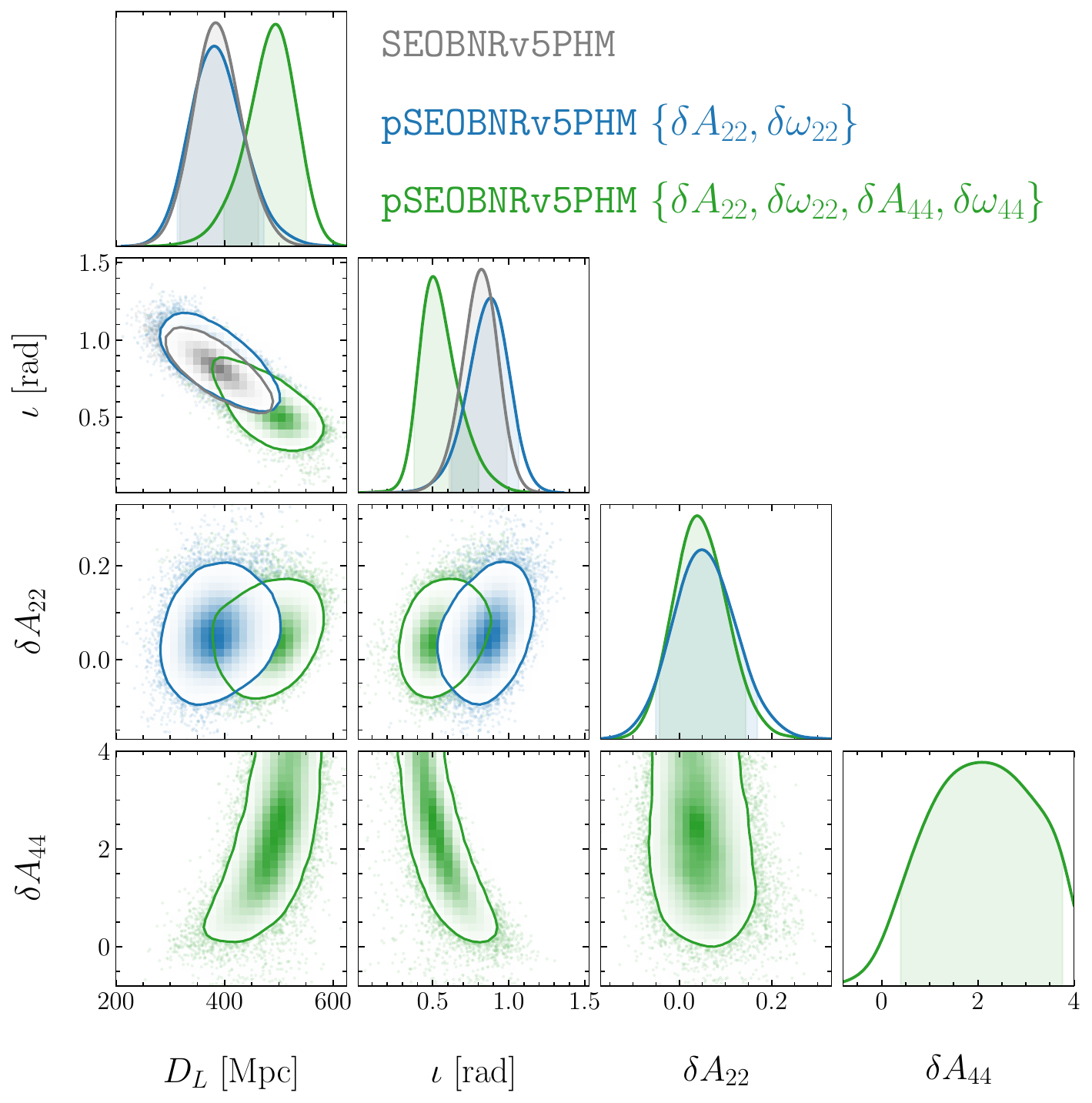}
    \caption{The one- and two-dimensional posterior distributions on the luminosity distance, $D_L$, the inclination angle, $\iota$ (folded to $[0, \pi/2]$), the $(2,2)$ mode amplitude deviation, $\delta A_{22}$ and the $(4,4)$ mode amplitude deviation, $\delta A_{44}$. The grey lines are obtained using as recovery model \texttt{SEOBNRv5PHM}, while the blue and green lines are obtained using \texttt{pSEOBNRv5PHM} with different deviation parameters. All contours indicate 90\% credible regions.}
    \label{correlations}
\end{figure}

Fig.~\ref{posteriors2} shows the posterior distribution on the merger amplitude deviation of the $(4,4)$ mode, which is marginally compatible with GR and rails against the upper bound of the prior. To better understand this result, we compare the intrinsic parameters of the binary obtained with this analysis to the parameter estimation of the event in GR~\cite{LIGOScientific:2025rid} to search for parameter correlations. 
As shown in Fig.~\ref{correlations}, we find that the posterior distribution on the luminosity distance is shifted with respect to the GR estimation, as well as the posterior distribution on the inclination angle. In Fig.~\ref{correlations} we show the posterior distributions from the analysis with \texttt{SEOBNRv5PHM} (grey lines) as compared to the ones from the analysis with parameter deviations $\{ \delta A_{22}, \delta \omega_{22}\}$ (blue lines) and $\{\delta A_{22}, \delta \omega_{22},\delta A_{44},\delta \omega_{44}\}$ (green lines). If we restrict $\delta A_{44}$ to the range $[-0.8,0.8]$, the shift of $D_L$ and $\iota$ from GR mitigates, confirming that it is associated with large $\delta A_{44}$ samples. Since the inclination angle is related to the relative amplitude of the $(2,2)$ and $(4,4)$ modes, a shift from the GR estimation can be compensated by the posterior on $\delta A_{44}$ shifting towards higher values to restore the correct $(2,2)$ to $(4,4)$ amplitude ratio. As a consequence, the luminosity distance changes in a way that the $(2,2)$ mode maintains the same overall amplitude.

\begin{figure}
    \centering
    \includegraphics[width=0.45\linewidth]{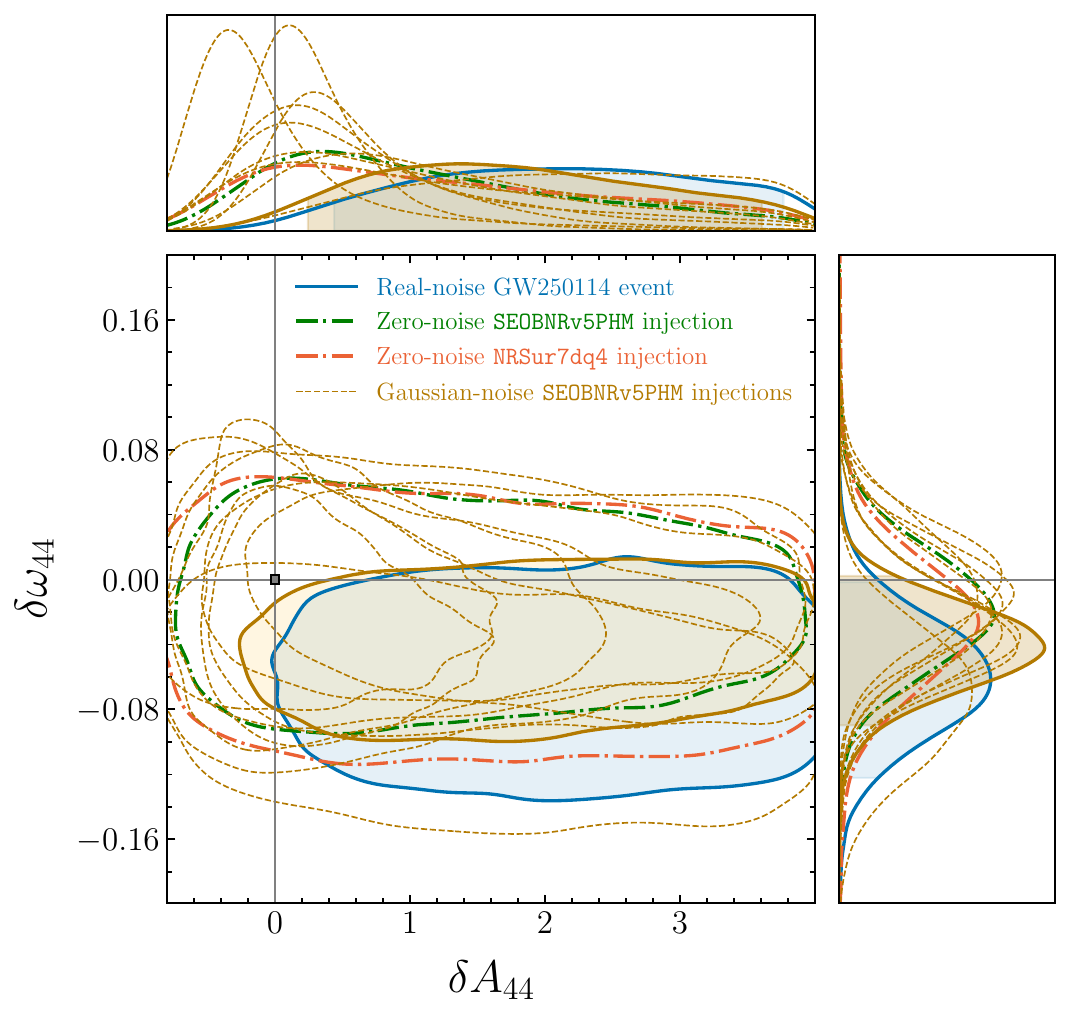}
    \caption{The one- and two-dimensional posterior distributions on the $(4,4)$ mode merger parameters for the \texttt{pSEOBNR} analysis of GW250114 with $\{\delta A_{22}, \delta \omega_{22}, \delta A_{44}, \delta \omega_{44}\}$ (blue curve) compared to the posterior distributions on the merger parameters for the same analysis with an \texttt{SEOBNRv5PHM} injection in zero-noise (green curve), \texttt{\texttt{NRSur7dq4}} injection in zero noise (red curve), and ten \texttt{SEOBNRv5PHM} injections in Gaussian noise (ocher curves). The specific realization in Gaussian noise that reproduces real-noise GW250114 analysis is highlighted with a thicker outline. All contours indicate 90\% credible regions and the vertical and horizontal lines mark the null GR expectation.}
    \label{44}
\end{figure}

To assess if this correlation explains the behavior of the posterior distribution on $\delta A_{44}$, we perform an injection in zero noise using the \texttt{SEOBNRv5PHM} waveform model. We find that the luminosity distance and the inclination angle are  shifted from the GR parameter estimation, proving that the correlation with $\delta A_{44}$ is genuine. The posterior distribution on $\delta A_{44}$ is wide and uninformative but, unlike the analysis on real data, peaks around the 0 and is compatible with GR. We rule out the possibility that waveform systematics is affecting the analysis by performing an injection in zero noise using the \texttt{NRSur7dq4} waveform model. Fig.~\ref{44} shows that the posterior distribution on the merger parameters of the $(4,4)$ mode is compatible between the \texttt{SEOBNRv5PHM} and \texttt{NRSur7dq4} injections.
Finally, we inject the \texttt{SEOBNRv5PHM} signal in ten different realizations of Gaussian noise, and find that some of them can induce the railing observed with the real data (see the thick ocher line in Fig.~\ref{44}). Indeed, one specific noise realization causes a shift away from GR for $\delta A_{44}$ and $\delta \omega_{44}$ that closely mirrors that of the analysis on real data. We conclude that the correlation between $\delta A_{44}$ and $\iota$ causes the posterior on the first parameter to have a long tail, while the specific noise realization of the event produces a shift of the distribution away from GR, causing it to rail against the upper bound.
The interplay between the correlation with the intrinsic parameters and the noise makes $\delta A_{44}$ particularly difficult to measure in this specific event.

\end{document}